\documentclass{PoS}
\usepackage{wrapfig}

\title{Kinetic and Chemical Equilibration of Quark-Gluon Plasma}
\ShortTitle{Kinetic and Chemical Equilibration of Quark-Gluon Plasma}

\author{\speaker{Xiaojian Du}\thanks{We thank Aleksi Kurkela, Aleksas Mazeliauskas, Jean-Francois Paquet, Ismail Soudi, Derek Teaney for valuable discussions. This work is supported by the Deutsche Forschungsgemeinschaft (DFG, German Research Foundation) – project number 315477589 – TRR 211. The authors also gratefully acknowledge computing time provided by the Paderborn Center for Parallel Computing (PC2) and National Energy Research Scientific Computing Center under US Department of Energy.}\\
        Fakultät für Physik, Universität Bielefeld,\\
  D-33615 Bielefeld, Germany\\
        E-mail: \email{xjdu@physik.uni-bielefeld.de}}

\author{Sören Schlichting\\
        Fakultät für Physik, Universität Bielefeld,\\
  D-33615 Bielefeld, Germany\\
  \email{sschlichting@physik.uni-bielefeld.de}}

\abstract{We solve a leading order QCD kinetic theory with light quarks and gluon degrees of freedom to study the non-equilibrium dynamics of the quark-gluon plasma (QGP). By including both elastic and inelastic scatterings for quarks and gluon, the model is proficient to describe kinetic and chemical equilibration of the QGP, and thus connects the initial (semi-) hard production of partons at early times with the hydrodynamic description of a near-thermalized quark-gluon plasma after the first fm/c of the collision. Within this approach, we investigate the time scales and mechanisms for kinetic and chemical equilibration of the QGP at zero and non-zero net-baryon density, and elaborate on the connections to jet quenching physics and hydrodynamics.}

\FullConference{%
  HardProbes2020\\
  1-6 June 2020\\
  Austin, Texas}

\begin{document}
\maketitle

\section{Introduction}
Exploration of collective phenomena within Quantum Chromo Dynamics (QCD) has been ongoing for decades based on studies from ultrarelativistic heavy-ion collisions (URHICs). The space-time dynamics of the near-equilibrium Quark-Gluon Plasma (QGP) produced in URHICs can be successfully described by relativistic viscous hydrodynamics with parameterized initial conditions at the end of the pre-equilibrium stage. Since the pre-equilibrium stage is expected to last only for $\sim 1 {\rm fm/c}$ after the collision, it has a comparatively small impact on the development of collective flow and other experimental observables. Nevertheless, a theoretical description of the pre-equilibrium dynamics is important from a conceptual point of view. While previous works have studied kinetic equilibration and onset of hydrodynamic behavior in pure Yang-Mills theory~\cite{Kurkela:2018vqr}, a dynamical description including quark and gluon degrees of freedom~\cite{Kurkela:2018oqw} is needed to capture the full glory of non-equilibrium QCD.
We establish a non-equilibrium QCD kinetic description including both quark and gluon degrees of freedom, to investigate kinetic and chemical equilibration of QCD plasmas with zero and finite charge/baryon density. Details of our study are provided in~\cite{Du:2020qcd}.

\section{QCD Kinetic Theory}
Our non-equilibrium simulations are based on an effective kinetic theory of QCD~\cite{Arnold:2002zm} including both elastic and inelastic scatterings between gluons and light quarks.
The corresponding Boltzmann equation describes the dynamics of species $a=g,u,\bar{u},d,\bar{d},s,\bar{s}$ at weak coupling
\begin{eqnarray}
\label{eq-bolzmann}
\frac{\partial}{\partial t}f_a(\vec{p},t)
=-C^{{2\leftrightarrow2}}_a[f](\vec{p},t)-C^{{1\leftrightarrow2}}_a[f](\vec{p},t)-C^{\rm exp}_a[f](\vec{p},t)
\end{eqnarray}
with collision integral of elastic scattering $C^{{2\leftrightarrow2}}_a[f]$, inelastic scattering $C^{{1\leftrightarrow2}}_a[f]$ and an additional longitudinal expansion term $C^{\rm exp}_a[f]$. Elastic $2\rightarrow 2$ scatterings include all leading order pQCD processes involving gluon and light quarks
\begin{eqnarray}
\label{eq-cint-elastic}
&&C^{{2\leftrightarrow2}}_a[f](\vec{p}_1)
=\frac{1}{2 \nu_{a}}\frac{1}{2 E_{p_1}}{\sum_{cd}}
\int d\Pi_{2\leftrightarrow2}
|\mathcal{M}_{cd}^{ab}(\vec{p}_1,\vec{p}_2|\vec{p}_3,\vec{p}_4)|^2F_{cd}^{ab}(\vec{p}_1,\vec{p}_2|\vec{p}_3,\vec{p}_4)
\end{eqnarray}
with statistical factor $F_{cd}^{ab}(\vec{p}_1,\vec{p}_2|\vec{p}_3,\vec{p}_4)$ fulfilling detailed balance, that eliminates the reactions in equilibrium, and scattering amplitudes $|\mathcal{M}_{cd}^{ab}(\vec{p}_1,\vec{p}_2|\vec{p}_3,\vec{p}_4)|$, which are screened by HTL constrained thermal masses~\cite{Kurkela:2018oqw}.
Effective $1\leftrightarrow 2$ inelastic processes describe the emission/absorption of quark/gluon radiation 
\begin{eqnarray}
\label{eq-ccint-inelastic-cast}
C^{{1\leftrightarrow2}}_a[f]
=\int_0^1\frac{dz}{2\nu_{a}}
\left[
\sum_{bc}\frac{d\Gamma_{bc}^{a}}{dz}\big(p,z\big)\nu_{a}F_{bc}^{a}(p|zp,\bar{z}p)-
\frac{2}{z^3}\frac{d\Gamma_{ab}^{c}}{dz}\big(\frac{p}{z},z\big)\nu_{c}F_{ab}^{c}(\frac{p}{z}|p,\frac{\bar{z}}{z}p)\right]
\end{eqnarray}
as described by an effective collinear rate $\frac{d\Gamma_{bc}^{a}}{dz}(p,z)$ (see e.g. \cite{Arnold:2002zm}), which encompasses the Bethe-Heitler regime at low energies as well as the Landau-Pomeranchuk Migdal regime at high energies. When considering systems which undergo a boost invariant longitudinal expansion, we employ co-moving $\tau,\eta$ coordinates, where the Boltzmann equation receives an additional contribution~\cite{Mueller:1999pi}
\begin{eqnarray}
\label{eq-cint-zexp}
C^{\rm exp}_a[f](\vec{p},\tau)
=-\frac{p_{z}}{\tau} \frac{\partial f_a(\vec{p},\tau)}{\partial_{p_{z}}}.
\end{eqnarray}

\section{Kinetic and Chemical Equilibration of the QGP}
Naturally, we shall first examine the equilibration of systems without longitudinal expansion, leading to a conservation of the total energy density and charge densities in the system.  Within this setup far-from-equilibrium systems can be characterized broadly into two categories, an over-occupied system dominated by a large number of low momentum gluons, or an under-occupied system with a small number of high-momentum gluon/quark mini-jets.
However, as a baseline examination, we first inspect the chemical equilibration of near-thermal systems. 

\textbf{Near-thermal Systems:} We prepare the system with either thermal gluons without quark/antiquark or thermal quark/antiquark pairs without gluons, such that the chemical equilibration of the system can be monitored in terms of the individual energy densities of gluons and quarks compared to their equilibrium values. We evaluate the time scale of chemical equilibration in terms of a kinetic relaxation time $\tau_R=\frac{4\pi\eta/s}{T_{\rm eq}}$
where the final equilibrium temperature $T_{\rm eq}$ can be estimated from the total energy density, and the shear viscosity over entropy density ratio depends on the strength of t'Hooft coupling $\lambda=g^2N_c$
\begin{figure}
	\begin{minipage}[b]{0.32\linewidth}
		\centering
		\includegraphics[width=1.10\textwidth]{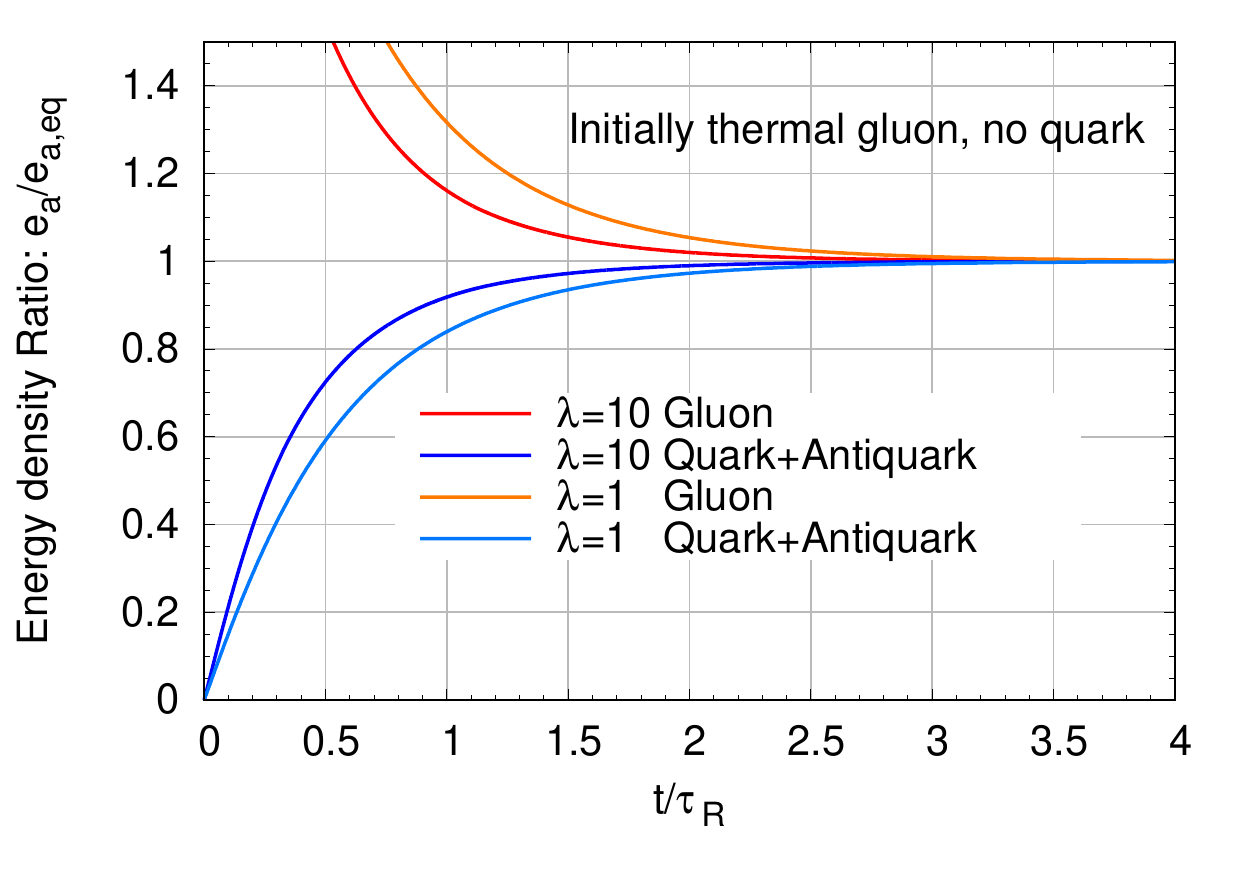}
	\end{minipage}
	\hspace{\fill}
	\begin{minipage}[b]{0.32\linewidth}
		\centering
		\includegraphics[width=1.10\textwidth]{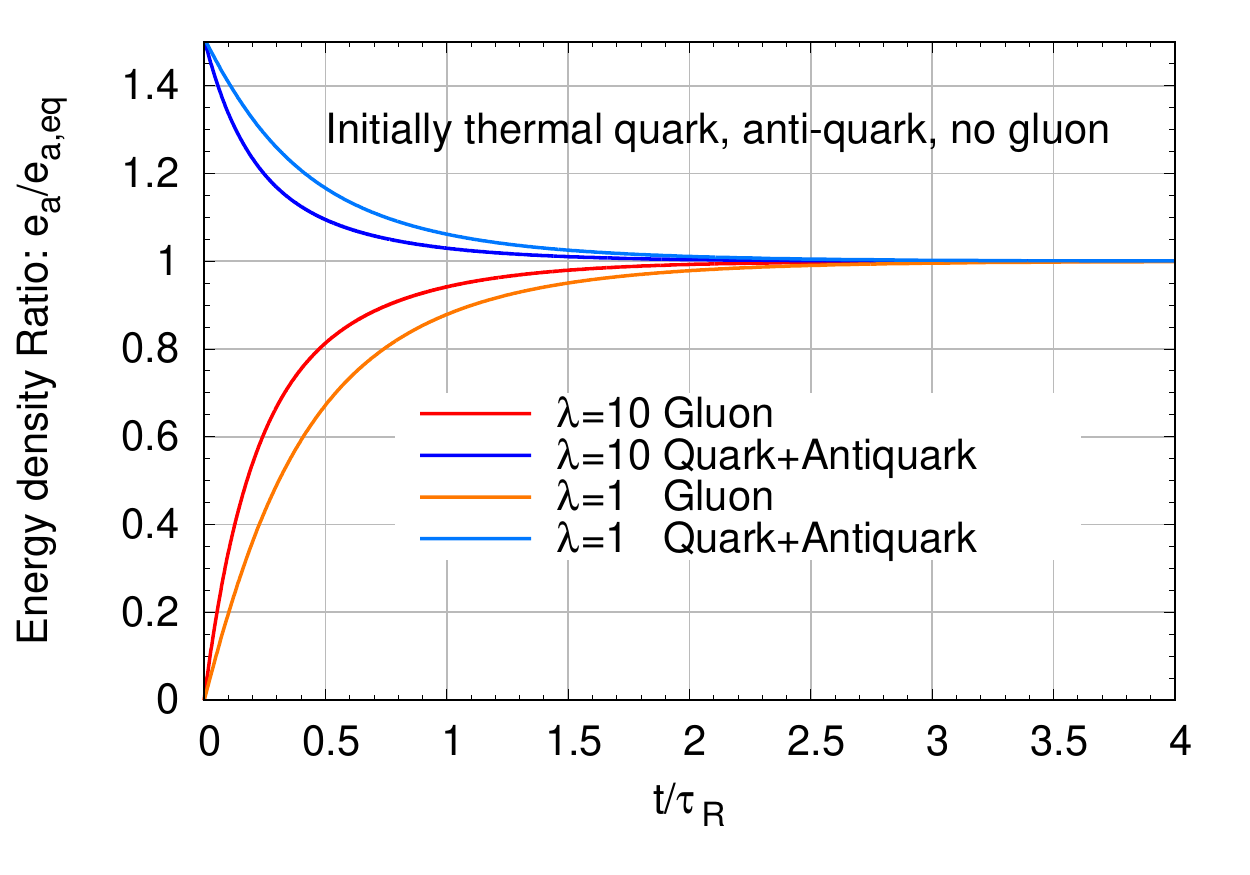}
	\end{minipage}
	\hspace{\fill}
	\begin{minipage}[b]{0.32\linewidth}
		\centering
		\includegraphics[width=1.10\textwidth]{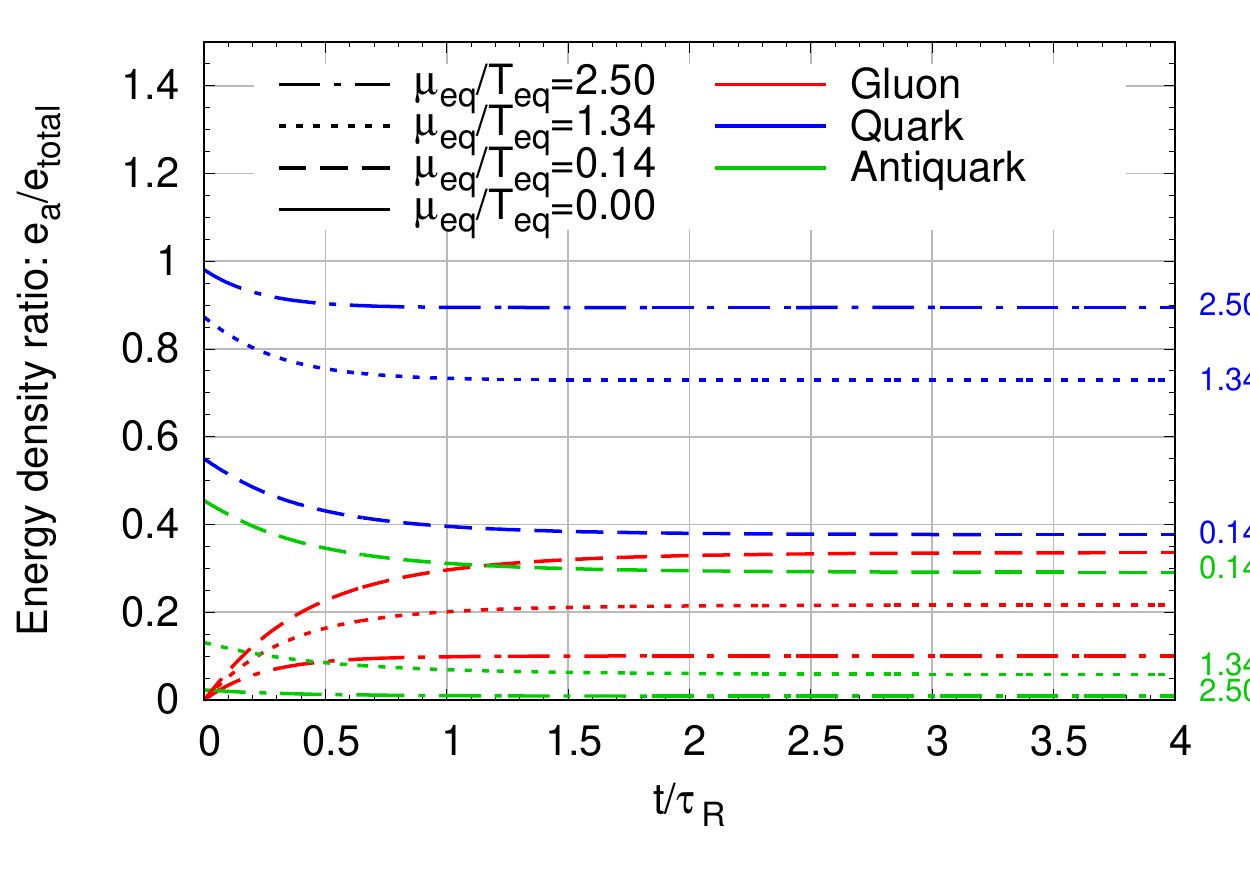}
	\end{minipage}
	\caption{Left and Middle panels: Evolution of the energy density of gluons and quarks plus antiquarks for different couplings at zero chemical potential. Right panel: Evolution of the energy density of gluons (red), quarks (blue) and antiquarks (green) at different chemical potentials $\mu_{eq}/T_{eq}=0 -- 2.5$ and coupling $\lambda$=1.}
	\label{fig-thermal}
\end{figure}
We present the evolution of the individual energy densities of gluons, quarks and antiquarks for different couplings $\lambda$=10 ($\eta/s\simeq 1$), $\lambda$=1 ($\eta/s\simeq 35$) in Fig.~(\ref{fig-thermal}). 
The energy densities of each species approach their equilibrium values at $\gtrsim 2\tau_R$, indicating a chemical equilibration at the same time scale as but slightly later than the kinetic equilibration.
Notably, with initially thermal gluon condition, the system equilibrates slightly slower, presumably caused by the ineffectiveness of quark/antiquark equilibration during the evolution.
The evolutions at different couplings show an overall universal merging curvature to their equilibrium values.
Evolutions of finite density systems in the right panel of Fig.~(\ref{fig-thermal}) show very similar results, with perhaps a slightly faster equilibration at finite density, due to the larger energy density in the presence of a non-vanishing chemical potential.

\begin{figure}
	\begin{minipage}[b]{0.32\linewidth}
		\centering
		\includegraphics[width=1.10\textwidth]{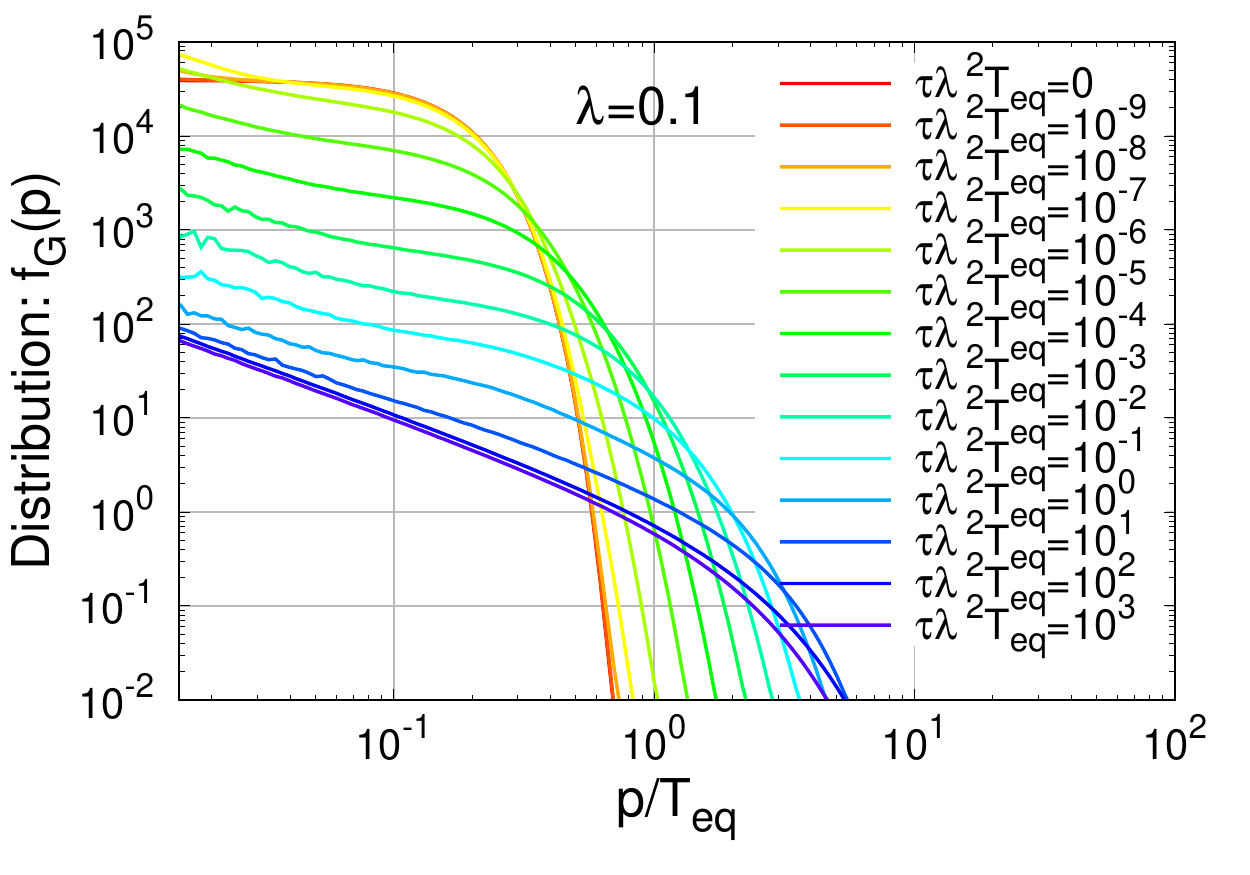}
	\end{minipage}
	\hspace{\fill}
	\begin{minipage}[b]{0.32\linewidth}
		\centering
		\includegraphics[width=1.10\textwidth]{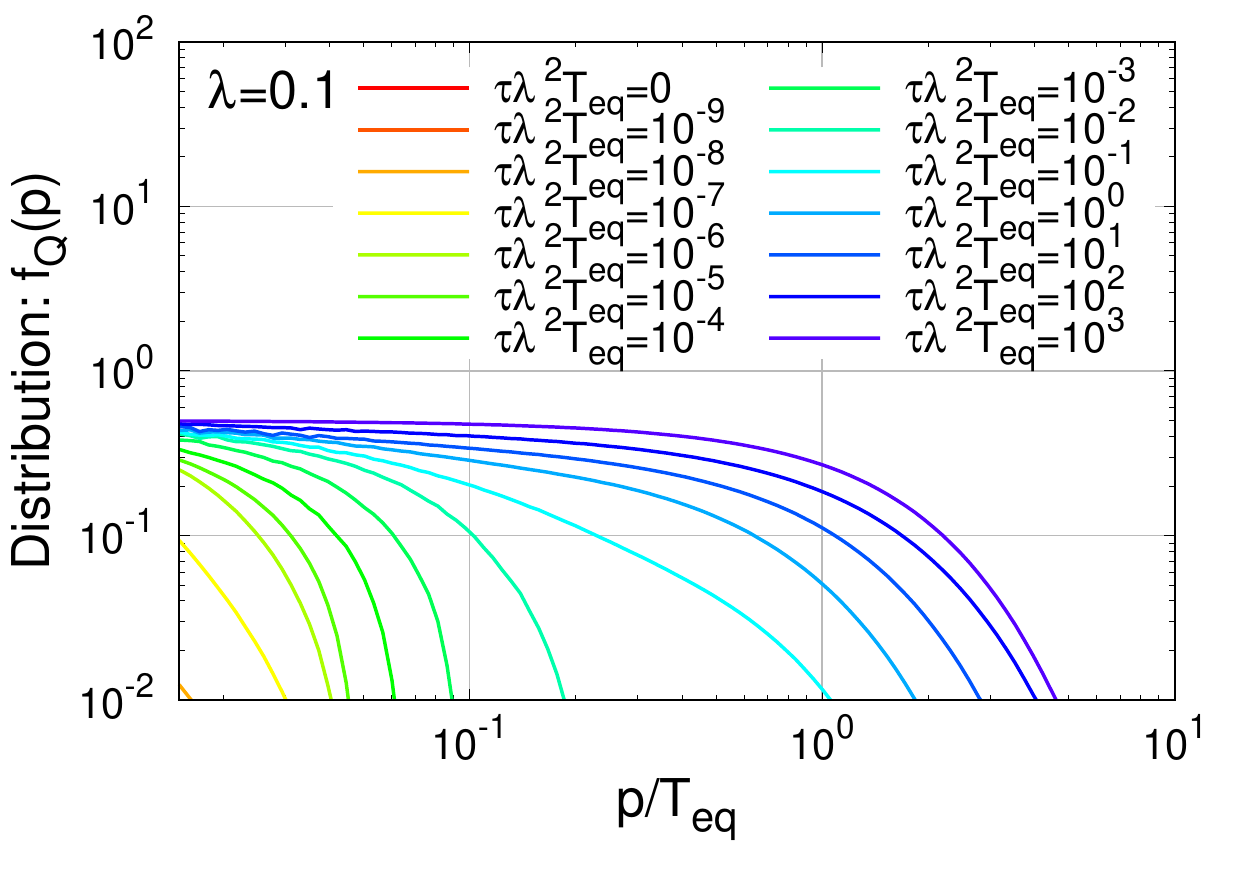}
	\end{minipage}
	\hspace{\fill}
	\begin{minipage}[b]{0.32\linewidth}
		\centering
		\includegraphics[width=1.10\textwidth]{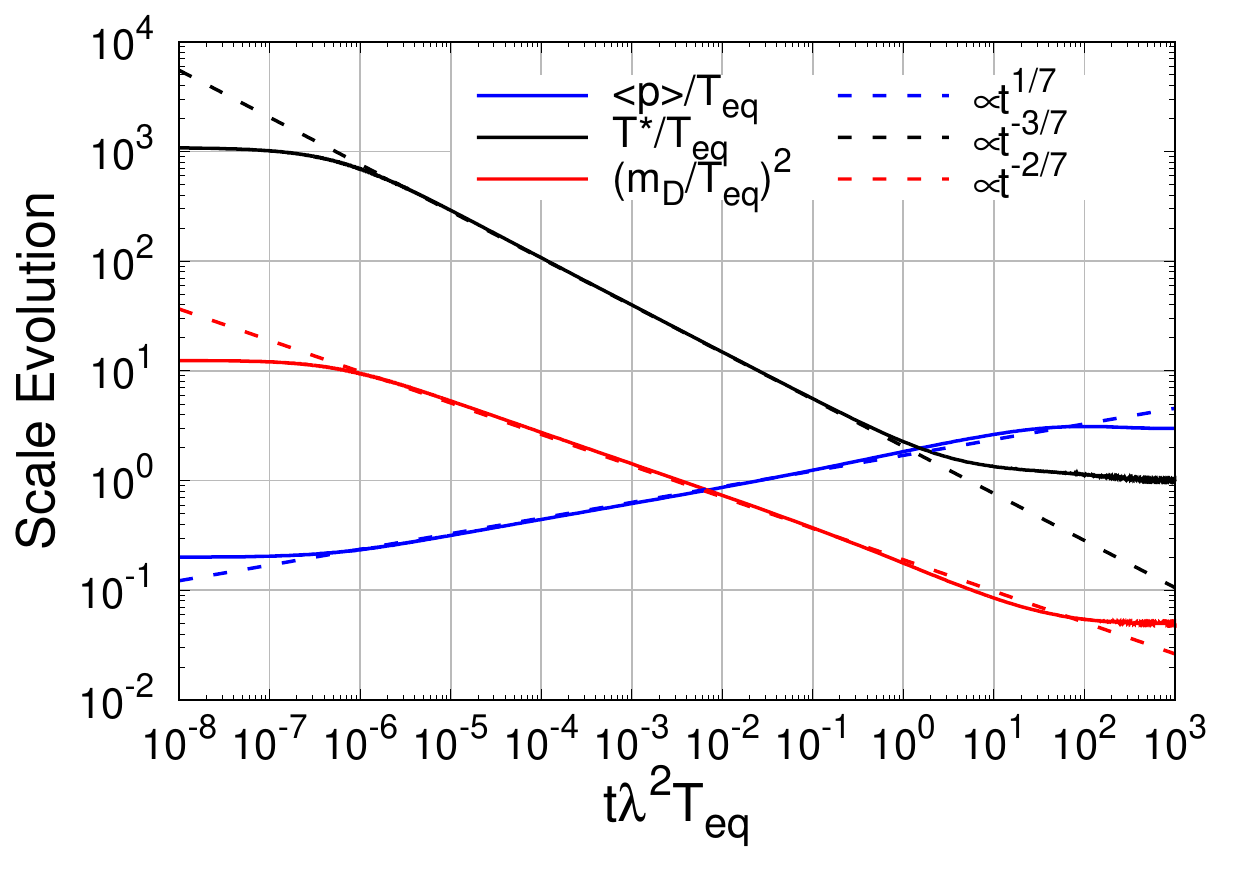}
	\end{minipage}
	\caption{Left and Middle panels: Gluon (left) and Quark ( right)spectra of an over-occupied system with initial average momentum $\left<p\right>_0/T{eq}=0.2$ at coupling $\lambda$=0.1. Right panel: The universal self-similar scalings of the same system.}
	\label{fig-overspectra}
\end{figure}

\textbf{Over-occupied Systems:} 
Over-occupied system are initially dominated by a large number of low momentum gluons, while the effects of quark/antiquarks are suppressed by the Pauli exclusion principle. The over-occupied gluon undergoes a quick memory loss of its initial spectra, followed by a self-similar direct energy cascade, before the system eventually approaches the equilibrium, as seen in Fig.~(\ref{fig-overspectra}). The spectra of quarks/antiquarks follow the gluon spectra, but the occupancies are limited due to Fermi statistics. Based on dimensional analysis of the underlying Boltzmann equation, screening masses $m_{D}^2,m_{q}^2$, effective temperature $T^{*}$  as well as the average momentum per particle $\left<p\right>$ are expected to follow a power-law evolution, which can be seen in the right panel of Fig.~(\ref{fig-overspectra}). Due to the dominance of gluons in the system, the power laws $m_D^2\sim t^{-2/7}$, $T^{*}\sim t^{-3/7}$, $\left<p\right>\sim t^{-1/7}$ predicted in \cite{Kurkela:2011ti} are not limited to pure-Yang Mills dynamics but also apply to QCD dynamics, and the same observations are also validated at larger couplings (see ~\cite{Du:2020qcd}).

\begin{wrapfigure}{l}{0.4\textwidth}
		\includegraphics[width=0.38\textwidth]{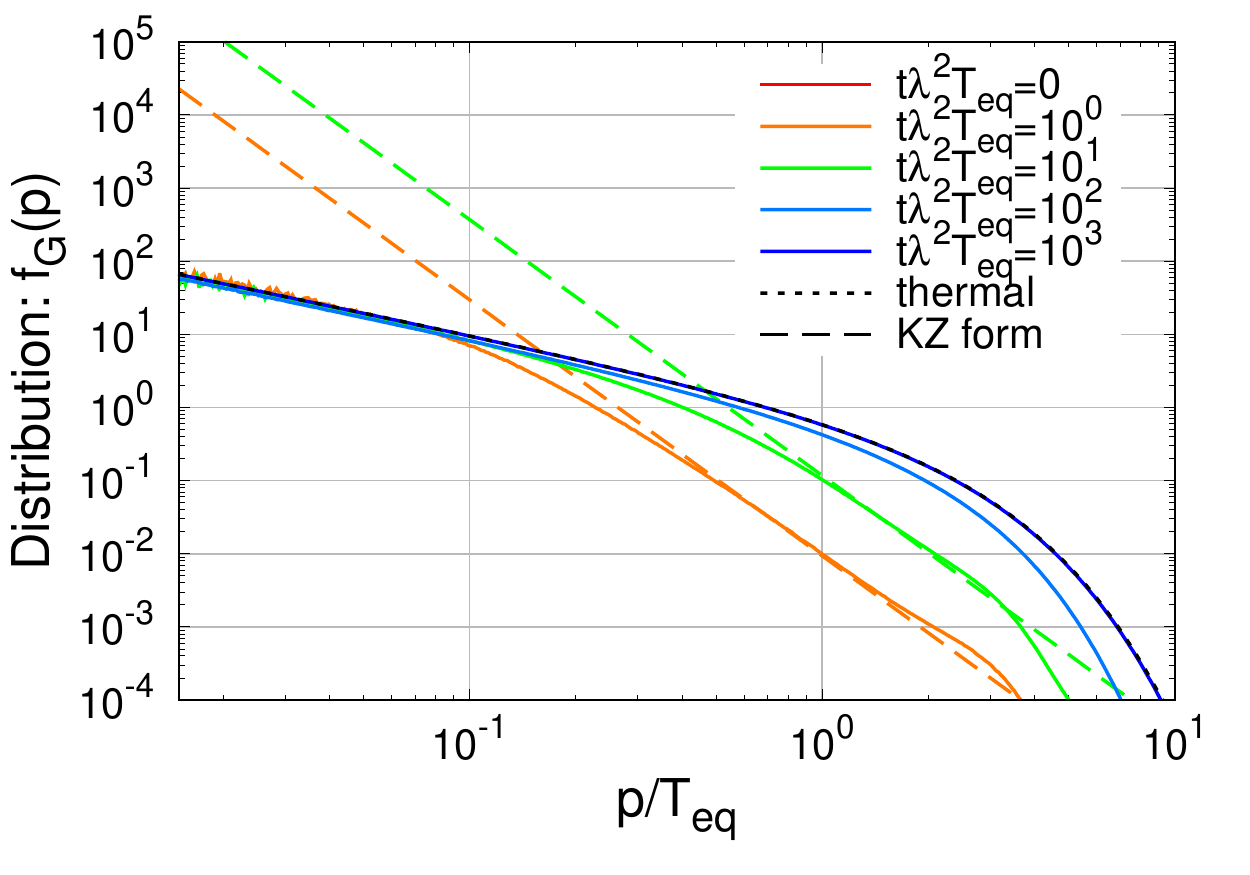}
	\hspace{\fill}
		\includegraphics[width=0.38\textwidth]{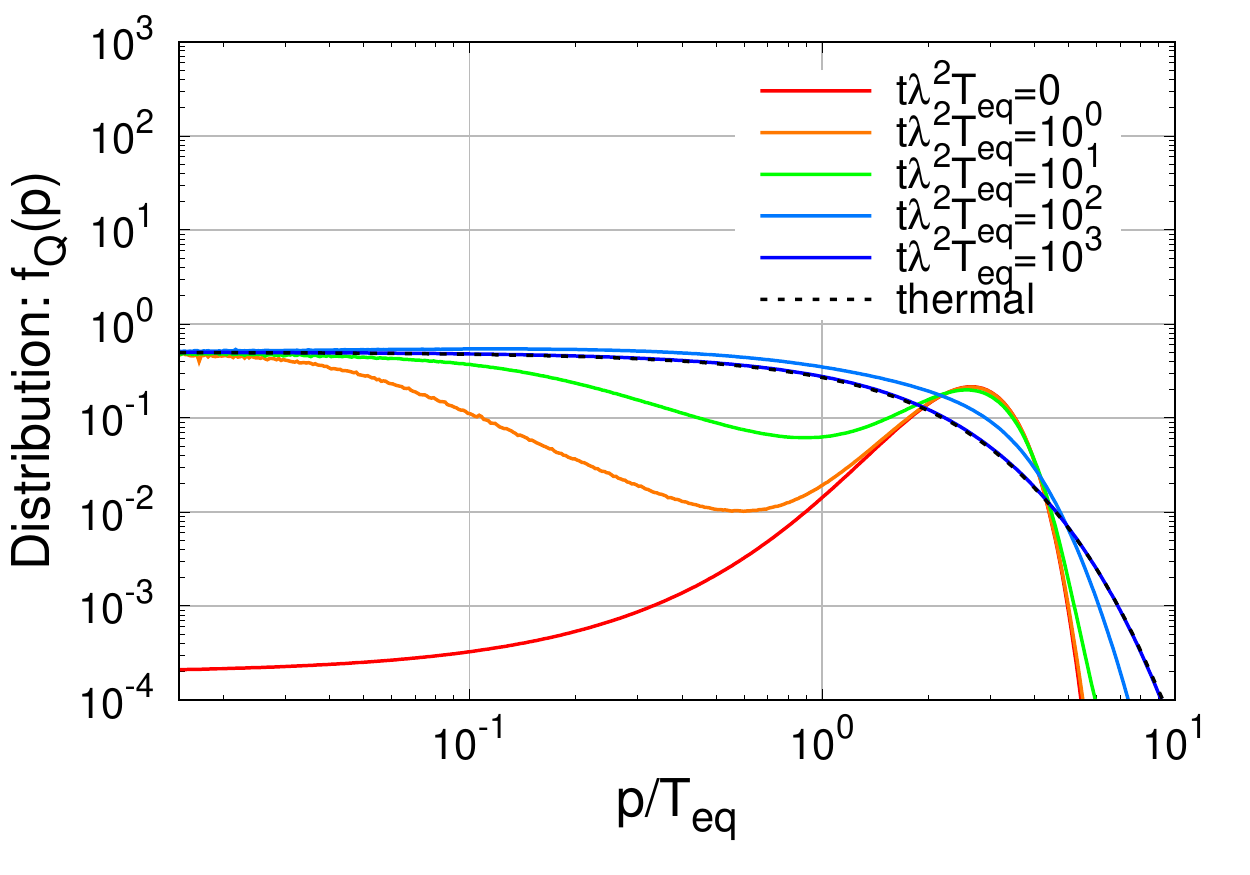}
	\caption{Evolution of gluon (top) and quark (bottom) spectra of an under-occupied system with initial separation of scale $\left<p\right>_0/T_{eq}=3$ at coupling $\lambda$=1.}
	\label{fig-underspectra}
\end{wrapfigure}

\textbf{Under-occupied Systems:} While over-occupied systems are dominated by low energy gluons, under-occupied systems are dominated by small number of high energy particles. Energy loss of hard particles and equilibration of the system are similar to mini-jet/jet quenching and equilibration in URHICs~\cite{Schlichting:2020lef}, undergoing the prominent "bottom-up thermalization" pattern~\cite{Baier:2000sb}. Fig.~(\ref{fig-underspectra}) shows spectra evolution of gluon and quark/antiquark for an initially under-occupied system of quark/antiquark pairs. Emission of primary quark/gluon radiation produces a soft thermal bath of quarks and gluons at low momenta. Subsequently, the remaining hard mini-jets loose energy to the soft bath. This inverse energy cascade is mediated by multi successive radiative branchings, and described by the Kolmogorov-Zakharov spectra~\cite{Mehtar-Tani:2018zba} at intermediate time scales. Eventually, the system thermalizes when the initial hard partons have lost all their energy to the soft thermal bath. More discussions with larger separations of scale and different mini-jets (gluon mini-jets/jets, quark mini-jets/jets at finite density) are presented in ~\cite{Du:2020qcd}.

\begin{wrapfigure}{l}{0.4\textwidth}
	\centering
	\includegraphics[width=0.38\textwidth]{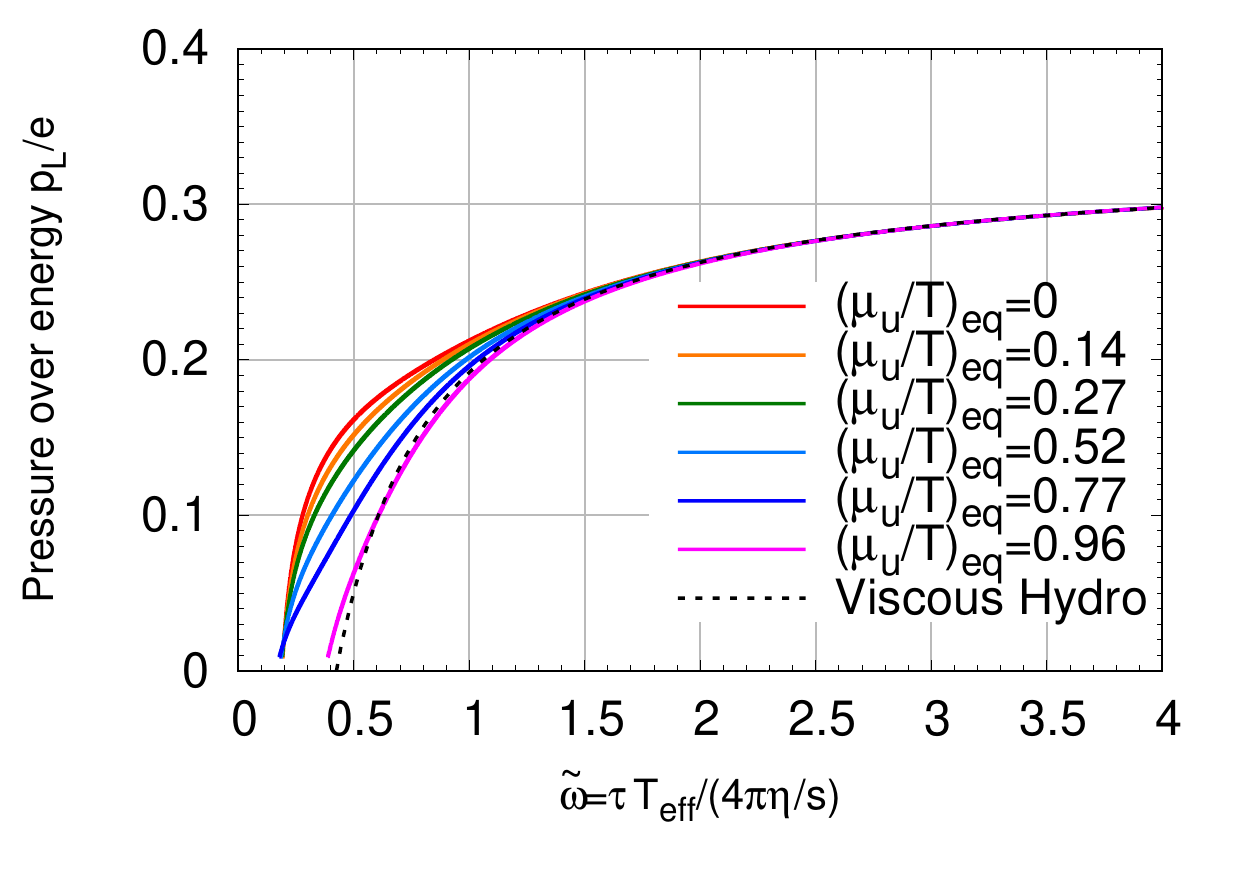}
	\hspace{\fill}
	\centering
	\includegraphics[width=0.38\textwidth]{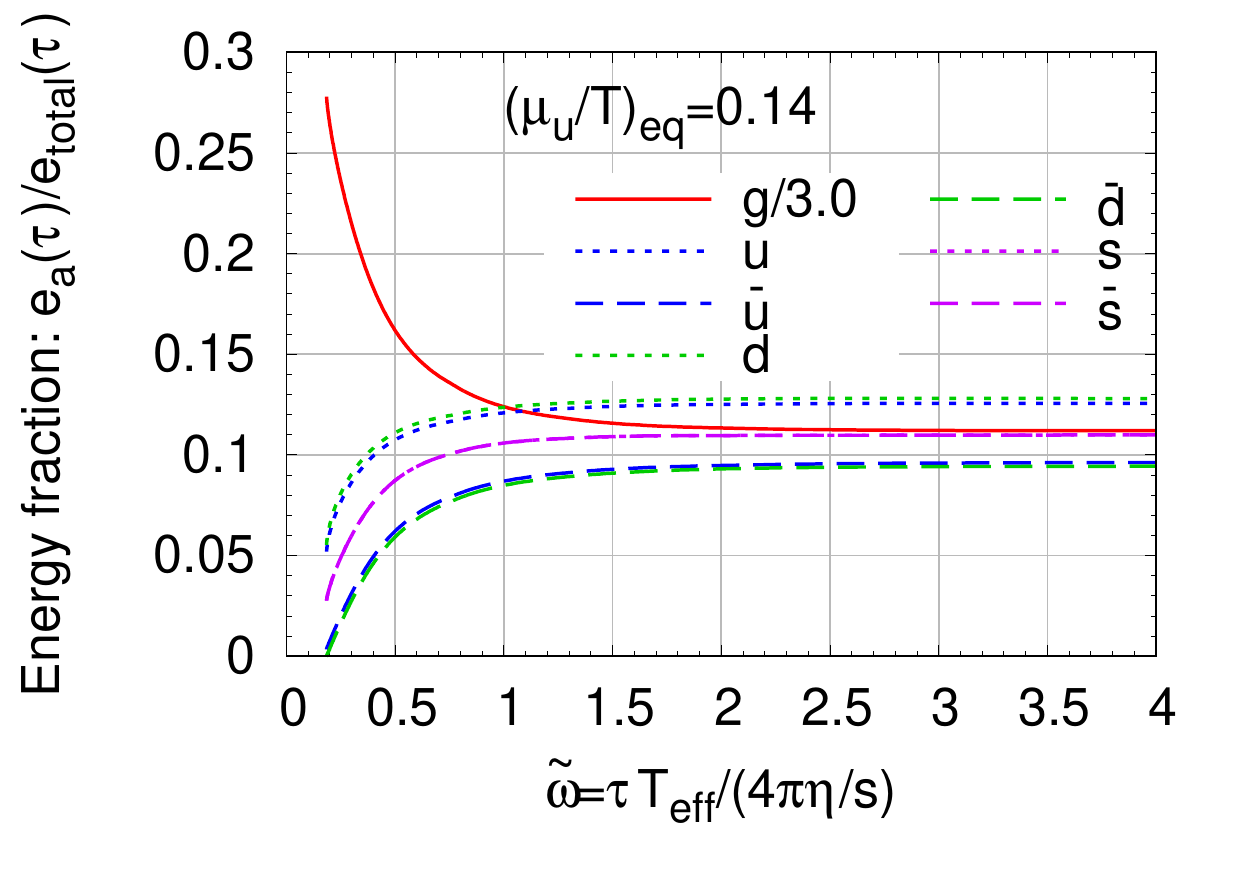}
	\caption{(top) Isotropization of expanding QGP for varying net-baryon density $\mu_{eq}/T_{eq}=0-1$. (bottom) Energy fractions of different species as an indicator of chemical equilibration.}
	\label{fig-expansion}
\end{wrapfigure}

\textbf{Longitudinal Expansion Systems:}
Now we move on with systems including expansion, most similar to the QGP produced in URHICs. In order to simulate a realistic pre-equilibrium dynamics we adopt gluon dominated initial conditions with zero and finite net-baryon density ($n_{Q}\simeq 0.4n_{B},n_{S}=0$). We summarize our main results in Fig~(\ref{fig-expansion}), where the top panel shows the evolution of the longitudinal pressure over energy densities $\frac{p_L}{e}$ at different densities compared to the first-order hydrodynamic asymptotes. We find that larger density systems exhibit slower isotropization, presumably due to ineffective quark contributions to isotropization. 
Chemical equilibration occurs on roughly the same time scale as kinetic equilibration, as can be inferred from the relative amount of energy carried by each species shown in the bottom panel of Fig~(\ref{fig-expansion}), where all species show an universal chemical equilibration time $\tilde{\omega} \simeq 2$ where $\tilde{\omega}=\tau T_{\rm eff}/(4\pi \eta/s)$ with $T_{\rm eff} \propto e^{1/4}$ is a dimensionless time scale for kinetic equilibration.

\section{Conclusions \& Outlook}
We developed a non-equilibrium QCD kinetic description to microscopically simulate the equilibration of the QGP far-from equilibrium. By including gluons as well as all light flavor quarks \& anti-quarks as explicit degrees of freedom, we are able to describe kinetic and chemical equilibration of the QGP at zero and non-zero densities of the conserved charges. Hence this framework provides a connection between the far-from equilibrium initial state and the initial conditions for hydrodynamic in URHICs. Beyond applications to URHICs, e.g. extending the pre-equilibrium description in K\o MP\o ST~\cite{Kurkela:2018vqr} to full QCD, this framework could also be used to access the QGP dynamics in the early universe.


\begin{thebibliography}{99}
	
\bibitem{Kurkela:2018vqr}
A.~Kurkela, et al,
Phys. Rev. C \textbf{99}, no.3, 034910 (2019)


\bibitem{Arnold:2002zm}
P.~B.~Arnold, G.~D.~Moore and L.~G.~Yaffe,
JHEP \textbf{01}, 030 (2003)
%
\bibitem{Kurkela:2018oqw}
A.~Kurkela and A.~Mazeliauskas,
Phys. Rev. D \textbf{99}, no.5, 054018 (2019)
%
\bibitem{Arnold:2002ja}
P.~B.~Arnold, G.~D.~Moore and L.~G.~Yaffe,
JHEP \textbf{06}, 030 (2002)
%
\bibitem{Mueller:1999pi}
A.~H.~Mueller,
Phys. Lett. B \textbf{475}, 220-224 (2000)
%
\bibitem{Du:2020qcd}
X.~Du and S.~Schlichting,
In preparation.
%
\bibitem{Schlichting:2020lef}
S.~Schlichting and I.~Soudi,
[arXiv:2008.04928].
%
\bibitem{Kurkela:2011ti}
A.~Kurkela and G.~D.~Moore,
JHEP \textbf{12}, 044 (2011)

\bibitem{Baier:2000sb}
R.~Baier, A.~H.~Mueller, D.~Schiff and D.~T.~Son,
Phys. Lett. B \textbf{502}, 51-58 (2001)

\bibitem{Mehtar-Tani:2018zba}
Y.~Mehtar-Tani and S.~Schlichting,
JHEP \textbf{09}, 144 (2018)
\end{thebibliography}
\end{document}